\documentclass[aps,twocolumn,preprintnumbers,amsmath,amssymb,floatfix]{revtex4}

\usepackage{graphicx}
\usepackage{longtable}
\usepackage{dcolumn}
\usepackage{bm}
\usepackage{color}
\usepackage[normalem]{ulem}
\usepackage[colorlinks=true, urlcolor=blue, linkcolor=blue, citecolor=blue]{hyperref}
\usepackage[all]{hypcap}
\newcommand{\ds}{\displaystyle}
\newcommand{\beq}{\begin{equation}}
\newcommand{\eeq}{\end{equation}}
\newcommand{\bea}{\begin{eqnarray}}
\newcommand{\eea}{\end{eqnarray}}

\begin{document}

\title{First-principles prediction of energy band gaps in 18-valence electron semiconducting 
half-Heusler compounds: Exploring the role of exchange and correlation}

\author{Emel G\"{u}rb\"{u}z$^{1}$}  
\author{Murat Tas$^{2}$}\email{murat.tas@gtu.edu.tr}
\author{Ersoy \c{S}a\c{s}{\i}o\u{g}lu$^{3}$}\email{ersoy.sasioglu@physik.uni-halle.de}
\author{Ingrid Mertig$^{3}$}
\author{Biplab Sanyal$^{1}$}
\author{Iosif Galanakis$^{4}$}\email{galanakis@upatras.gr}

\affiliation{$^{1}$Department of Physics and Astronomy, Uppsala University, 75120 Uppsala, Sweden \\
$^{2}$Department of Physics, Gebze Technical University, 41400 Kocaeli, Turkey \\
$^{3}$Institute of Physics, Martin Luther University Halle-Wittenberg, 06120 Halle (Saale), Germany \\
$^{4}$Department of Materials Science, School of Natural Sciences, University of Patras, GR-26504 
Patra, Greece}

\date{\today}

\begin{abstract}

The choice of exchange functional is a critical factor in determining the energy band gap of 
semiconductors. \textit{Ab initio} calculations using different exchange functionals, 
including the conventional generalized-gradient approximation (GGA) functionals, meta-GGA 
functionals, and hybrid functionals, show significant differences in the calculated energy 
band gap for semiconducting half-Heusler compounds. These compounds, which have 18 valence 
electrons per unit cell, are of great interest due to their thermoelectric properties, 
making them suitable for energy conversion applications. In addition, accounting for 
electronic correlations using the $GW$ method also affects the calculated energy band gaps 
compared to standard GGA calculations. The variations in calculated energy band gaps are 
specific to each material when using different functionals. Hence, a detailed investigation 
of the electronic properties of each compound is necessary to determine the most appropriate 
functional for an accurate description of the electronic properties. Our results indicate 
that no general rules can be established, and a comparison with experimental results is 
required to determine the most appropriate functional.

\end{abstract}

\maketitle

\section{\label{sec1} Introduction}

In 1964, Hohenberg and Kohn established the fundamental theorems of density functional theory 
(DFT), which state that the total energy of a system can be expressed uniquely as a functional 
of the electronic charge density \cite{Hohenberg1964}. The following year, Kohn and Sham made 
significant progress by formulating a quantum mechanical treatment of the many-electron problem 
as an equivalent single-electron problem, where each electron interacts with the mean-field 
generated by the other electrons \cite{Kohn1965}. This seminal work forms the basis of modern 
\textit{ab-initio} (first-principles) electronic band structure calculations, which solve the 
Kohn-Sham equations through self-consistent algorithms. An essential component of this 
transition to the single-electron Kohn-Sham equations is the inclusion of the 
exchange-correlation energy ($E_\mathrm{xc}$). The exchange energy ($E_\mathrm{x}$) arises 
from the indistinguishability of electrons as fermions. As an electron moves within the 
mean-field produced by the other electrons, it locally alters the electronic charge 
distribution, a phenomenon described by the correlation energy ($E_\mathrm{c}$). Unfortunately, 
exact forms of neither the exchange nor the correlation energies are known, necessitating the 
utilization of approximations to describe them.

In 2001, Perdew and Schmidt introduced the concept of Jacob's ladder for the exchange energy 
($E_\mathrm{x}$) \cite{Perdew01}. As we move up this ladder,
the exchange functionals are more complicated to build and use but 
potentially more accurate. The ladder starts with the local-density functional 
approximation (LDA), which expresses $E_\mathrm{x}$ as a functional of the electronic charge 
density ($n(\mathbf{r})$), assuming a uniform electron gas. However, LDA is known to be 
inadequate in accurately describing the ground state of certain elements (\textit{e.g.}, Fe) 
and significantly underestimates the lattice constants of materials. Moving up the ladder, 
we encounter the Generalized Gradient Approximation (GGA) functionals, where $E_\mathrm{x}$ is 
expressed in terms of both the electronic charge density ($n(\mathbf{r})$) and its gradient 
($\nabla n(\mathbf{r})$) \cite{Perdew1991}. The GGA was more successful than LDA, as the 
inclusion of the gradient provides a simplified account of the non-local nature of the charge 
density. The Perdew-Burke-Ernzerhof (PBE) form of GGA, developed in 1996, is widely used 
\cite{Perdew1996}. While GGA functionals accurately capture the structural properties of most 
semiconductors, they struggle to reproduce energy band gaps with high precision. Consequently, 
in order to address this limitation, one typically needs to progress to the higher levels of 
Jacob's ladder, specifically the third or fourth steps.

The third stair contains the so-called meta-GGA functionals where $E_\mathrm{x}$ is also a 
function of the Laplacian of the charge density $\nabla^2 n(\textbf{r})$ \cite{Perdew01}. 
The most recent form of these functionals is the non-empirical strongly constrained and 
appropriately normed (SCAN) meta-GGA functional which fulfills all known constraints that 
the exact density functional must fulfill, and improves significantly over LDA and PBE the 
geometries and energies of diversely-bonded materials (including covalent, metallic, ionic, 
hydrogen, and van der Waals bonds) \cite{SCAN}. The most recent version of SCAN is known as 
r$^2$SCAN (second revision of SCAN) and was developed by Furness and collaborators in 2020 
\cite{r2SCAN}. The fourth stair contains the so-called hybrid functionals where the 
$E_\mathrm{x}$ is written as a mixture of the exact Hartree exchange energy and of the GGA 
or meta-GGA exchange \cite{Becke1993}. The most widely used hybrid functional for crystals 
is the so-called PBE0 where the coefficients of the Hartree and GGA exchange energies are 
$\frac{1}{4}$ and $\frac{3}{4}$ respectively \cite{PBE0}. In 2003 Heyd and collaborators 
altered the PBE0 hybrid functional using an error-function-screened Coulomb potential to 
calculate the exchange portion of the energy in order to improve computational efficiency, 
especially for metallic systems; this functional is known as HSE03 (Heyd–Scuseria–Ernzerhof) 
\cite{HSE03}. In 2006 the same authors changed the screening parameters to better reproduce 
experimental results and the modified functional is known as HSE06 \cite{HSE06}. The drawback 
of these hybrid functionals (PBE0, HSE03, HSE06) is the predefined coefficients of the 
Hartree and GGA energies which are not optimal for all semiconductors and in many cases 
produce very large energy band gaps. A special case of functionals is the Becke-Johnson (BJ) 
functionals developed in 2006 \cite{BJ}. It is categorized as a meta-GGA functional and it 
is a parameter-dependent empirical functional aiming to reproduce, as accurately as hybrid 
functionals, the energy band gaps of semiconductors but in a much more efficient way. In its 
initial form, the total energy of the studied material is not variant with respect to the BJ 
functional. It became very popular in 2009, when Tran and Blaha modified it so that the total 
energy is variational with respect to it and incorporated it in first-principles electronic 
bands structure methods; the new functional in literature is called modified Becke Johnson 
functional (mBJ) or Tran-Blaha 2009 functional (TB09) \cite{TB09}. Although it is widely 
used in modern electronic band structure calculations, its accuracy is under question since 
the parameters which it contains have been optimized to fit the energy band gaps of certain 
semiconductors and are not universal \cite{TB09b,Meinert2013}.

All functionals mentioned above concern only the exchange part of the exchange-correlation 
functional. The correlation energy $E_\mathrm{c}$ is much smaller in magnitude than the 
$E_\mathrm{x}$ but plays a crucial role in many physical properties. The first attempt to 
describe $E_\mathrm{c}$ accurately was made by Vosko, Wilk and Nusair in 1980 (VWN80) 
\cite{VWN80}, who fitted a functional similar to the LDA for the $E_\mathrm{x}$ to the 
Quantum Monte Carlo results of Ceperley and Alder \cite{CA}. In 1992 Perdew and Wang 
presented a new functional for the correlation energy (PW92) which is widely used in all 
electronic band structure calculations in conjunction with all LDA, GGA, meta-GGA, and 
hybrid exchange functionals \cite{PW92}. There are two common ways to include electronic 
correlations in the first-principles electronic structure calculations. The first one is 
the so-called LDA+$U$ or GGA+$U$ scheme where an effective on-site Coulomb repulsion term 
Hubbard $U$ and Hund exchange $J$ are used to account for the correlation effects 
\cite{LDAU1,LDAU2}. A more elaborate modern computational scheme, which has resulted from 
the merging of the DFT and many-body Hamiltonian methods are the so-called LDA(GGA)+DMFT 
where DMFT stands for dynamical mean field theory \cite{DMFT1,DMFT2}. Although both 
approaches are successful in some cases, they suffer from the need to know in advance the 
$U$ and $J$ parameters which are materials-specific. These parameters can be difficult to 
determine experimentally, and their computational calculation is also a tedious task 
\cite{UJ1,UJ2}. A correct treatment of the correlation energy involves description of the 
electronic excitation spectrum which is not correctly represented by the solution of the 
Kohn-Sham equations; this is not a failure of the correlation functionals but rather 
because the Kohn–Sham eigenvalues are not meant to be interpreted as the excitation 
energies of the real interacting system. A more elaborated method for the correct 
description of the excited states is the \textit{GW} approximation for the electronic 
self-energy, which is derived in the framework of many-body perturbation theory and, thus, 
treats the interactions among the electrons beyond the mean-field approximation \cite{GW}. 
This means that the $GW$ approximation can account for a larger part of electronic 
correlation than DFT, which can have a strong effect on the band gaps of semiconductors 
and insulators \cite{GW2}. The $GW$ approximation has been shown to be quite successful in 
describing the excited states of materials, including the band gaps and optical properties. 
It is particularly useful for materials with weak to intermediate correlations, where DFT 
often fails to give accurate results \cite{GW3}.

\begin{figure}[htbp]
\begin{center}
\includegraphics[width=0.48\textwidth]{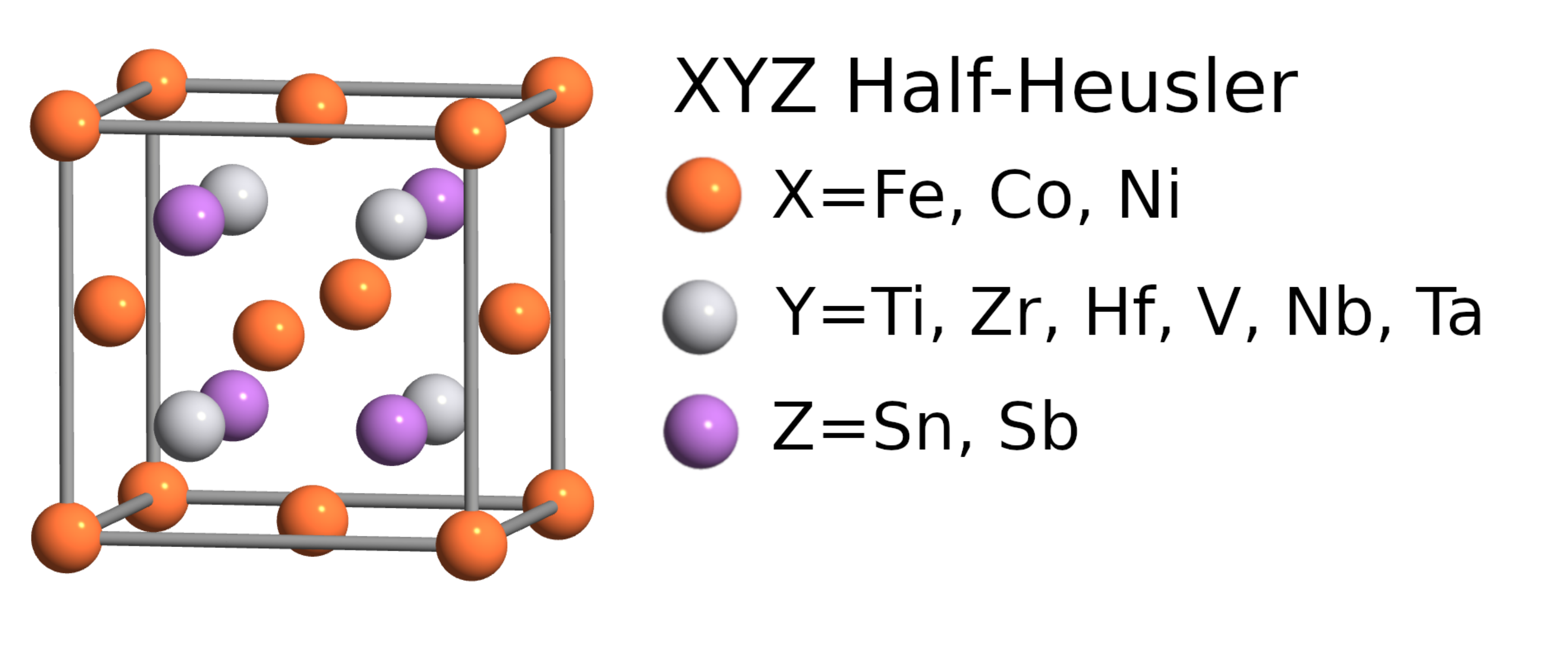}
\end{center}
\vspace*{-0.6cm}
\caption{(Color online) Schematic representation of the $C1_\mathrm{b}$ lattice structure 
adopted by the half-Heusler compounds.} \label{fig1}
\end{figure}

\section{\label{sec2}Heusler compounds and motivation}

Heusler compounds, named after Fritz Heusler \cite{Heusler1903,Heusler1912}, are ternary and 
quaternary intermetallic compounds that crystallize in close-packed lattice structures 
\cite{Graf2011,Tavares2023,Chatterjee2022}. Among them the ternary compounds having the 
chemical formula $XYZ$, where $X$ and $Y$ are transition-metal atoms and $Z$ is a metalloid, 
are named half-Heusler or semi-Heusler compounds. When the number of valence electrons in the 
unit cell exceeds 19 and goes up to 22, most of them are half-metallic metals, and the total 
spin magnetic moment in the unit cell in units of $\mu_\mathrm{B}$ follows the Slater-Pauling 
rule $M_\mathrm{t}=Z_\mathrm{t}-18$, where $Z_\mathrm{t}$ is the total number of valence 
electrons in the unit cell \cite{Galanakis2002a,Galanakis2023}. The number 18 expresses the 
fact that there are exactly nine occupied states in the minority-spin electronic band 
structure, which exhibits semiconducting behavior. There is a single $s$ and a triple $p$ 
band low in energy stemming from the $Z$ atom. The $d$ valence orbitals of the $X$ and $Y$ 
atoms hybridize creating five occupied bonding orbitals, which are separated by an energy gap 
from the five unoccupied antibonding orbitals. The Slater-Pauling rule correctly predicts that 
half-Heusler compounds with exactly 18 valence electrons should be non-magnetic semiconductors 
with a gap in both spin-channels \cite{Galanakis2002a,Galanakis2023}. This ‘‘18-electron rule" 
for semiconducting half-Heusler compounds was also derived by Jung \textit{et al.} based on 
ionic arguments \cite{Jung2000}. Among the 18-valence electron half-Heusler compounds, CoTiSb, 
NiTiSn, FeVSb, and CoVSn have attracted most of the attention. Pierre and collaborators in 
1994 have confirmed experimentally the non-magnetic semiconducting character of NiTiSn 
\cite{PIERRE1994}. Tobola \textit{et al.} have shown experimentally that CoTiSb is also a 
non-magnetic semiconductor \cite{Tobola1998}. The experimental findings for both NiTiSn and 
CoTiSb have been also confirmed by \text{ab-initio} calculations in 
Refs. \cite{Tobola1998,Tobola2000}. Recently, Ouardi \textit{et al.} have synthesized CoTiSb 
and investigated it both theoretically and experimentally \cite{Ouardi2012}. Lue and 
collaborators grew samples of CoVSn and their findings were consistent with a non-magnetic 
semiconducting behavior \cite{Lue2001}. Finally, Mokhtari and collaborators have shown 
theoretically that FeVSb is also a non-magnetic semiconductor \cite{Mokhtari2018} followed 
by the experimental observation in 2020 by Shourov \textit{et al.} \cite{Shourov2020}. 
Ma \textit{et al.} in 2017 studied using first-principles calculations a total of 378 
half-Heusler compounds \cite{Ma2017}. Among them, there were 27 compounds with 18 valence 
electrons, including the aforementioned ones, which were all found to be non-magnetic 
semiconductors \cite{Ma2017}.

Semiconducting half-Heusler compounds are of particular interest due to their potential 
applications in thermoelectric and optical devices \cite{Tavares2023}. Moreover, their 
implementation in low-dimensional devices can lead to exotic behaviors, such as spin-polarized 
electron/hole gas, which are of interest for spintronic applications \cite{Emel2023}. For all 
of these applications, correct description of their electronic band structure is essential. 
Since experimental data as mentioned above are limited to the verification of semiconducting 
character without going into details of the electronic band structure, it is important to 
investigate suitability of various functionals to accurately describe the band structure of 
these 18 valence electron semiconducting Heusler compounds. This study focuses on 12 such 
Heuslers: Fe(V,Nb,Ta)Sb, Co(Ti,Zr,Hf)Sb, Co(V,Nb,Ta)Sn and Ni(Ti,Zr,Hf)Sn. We employ all PBE, 
mBJ, r$^2$SCAN, PBE0, and HSE06 functionals for the exchange energy as well as the $GW$ 
method to correctly describe the correlation energy. We discuss the resulting electronic 
band structure in all cases. The rest of the manuscript is organized as follows: In Section 
\ref{sec3} we give details of our calculations, in Sections \ref{sec4} and \ref{sec5} we 
present our results, and finally, in Section \ref{sec6} we summarize our results and present 
the conclusions of our study.

\section{\label{sec3}Computational details}

Half-Heusler compounds crystallize in the $C1_\mathrm{B}$ lattice structure shown in Fig. 
\ref{fig1}. The conventional unit cell is the cube shown in the figure, while the primitive 
unit cell containing exactly a formula unit (f.u.) 
is a fcc one with three atoms as basis $X$ at $(0\,0\,0)$, $Y$ at $(\frac{1}{4}\,
\frac{1}{4}\,\frac{1}{4})$, and $Z$ at $(\frac{3}{4}\,\frac{3}{4}\,\frac{3}{4})$ in Wyckoff 
coordinates. We adopted the lattice constants calculated in The Open Quantum Materials 
Database (OQMD) for all twelve materials \cite{oqmd}, and we present them in the first 
column in Table \ref{table1}. The values are between 5.78 and 6.15 \AA\ being comparable to 
the values of cubic binary semiconductors. Our tests show that the lattice constants 
presented in OQMD, where the PBE functional has been used, differ less than 1 \%\ 
from the PBE equilibrium ones calculated with the electronic band structure methods 
employed in the current study. In Table \ref{table1} we also present the 
formation energy and the hull distance calculated in OQMD. Not only formation energies are 
negative meaning that the creation of the crystal is favored, but also the hull distance is 
exactly zero in all cases (with the exception of CoVSn) which means that the $C1_\mathrm{B}$ 
lattice is the most stable phase. For CoVSn, the hull distance has a value of 0.01 eV/atom 
which is very small and one can easily argue that when grown at finite temperature it will 
form a $C1_\mathrm{B}$ crystal and will not decompose in a mixture of simpler phases.

\begin{table*}[t]
\caption{\label{table1}
Lattice constants $a_0$, formation energy ($E_\text{form}$), 
convex hull distance energy ($\Delta E_\text{con}$), and calculated band gap values with 
one-shot $GW$, and PBE, mBJ, r$^2$SCAN, HSE06 and PBE0 functionals for all half-Heusler 
compounds studied. The $a_0$, $\Delta E_\text{con}$, and $E_\text{form}$ values are taken 
from The Open Quantum Materials Database \cite{oqmd}.} 
\begin{ruledtabular}
\begin{tabular}{@{}l*{10}{c}@{}}
Compound & $a_0$  & $E_\text{form}$ & $\Delta E_\text{con}$ & 
\multicolumn{7}{c}{$E_{\mathrm{g}}$ (eV)} \\
\cline{5-11}
& ({\AA}) &(eV/at.) & (eV/at.) & \multicolumn{2}{c}{FLEUR} & 
\multicolumn{5}{c}{\textsc{QuantumATK}} \\
\cline{5-6} \cline{7-11}
       &      &       &      & $GW$ & PBE  & PBE  & mBJ  & r$^2$SCAN & HSE06 & PBE0 
\\ \hline
FeVSb  & 5.78 & -0.21 & 0.00 & 0.51 & 0.34 & 0.34 & 0.78 & 0.57 & 1.36 & 2.04  \\
FeNbSb & 5.96 & -0.35 & 0.00 & 0.32 & 0.53 & 0.52 & 0.61 & 0.70 & 1.22 & 1.87  \\
FeTaSb & 5.95 & -0.30 & 0.00 & 0.71 & 0.86 & 0.86 & 0.90 & 1.04 & 1.49 & 2.09  \\
CoTiSb & 5.88 & -0.69 & 0.00 & 1.34 & 1.07 & 1.09 & 1.03 & 1.19 & 1.42 & 2.09  \\
CoZrSb & 6.09 & -0.76 & 0.00 & 1.21 & 1.07 & 1.07 & 1.09 & 1.20 & 1.56 & 2.15  \\
CoHfSb & 6.05 & -0.72 & 0.00 & 1.32 & 1.13 & 1.14 & 1.13 & 1.29 & 1.51 & 2.13  \\
CoVSn  & 5.79 & -0.17 & 0.01 & 0.73 & 0.64 & 0.63 & 0.88 & 0.84 & 1.32 & 2.04  \\
CoNbSn & 5.96 & -0.36 & 0.00 & 0.82 & 1.00 & 1.00 & 0.94 & 1.15 & 1.47 & 2.12  \\
CoTaSn & 5.95 & -0.32 & 0.00 & 1.19 & 1.05 & 1.06 & 1.07 & 1.21 & 1.51 & 2.22  \\
NiTiSn & 5.93 & -0.58 & 0.00 & 0.67 & 0.46 & 0.46 & 0.39 & 0.53 & 0.59 & 1.23  \\
NiZrSn & 6.15 & -0.72 & 0.00 & 1.01 & 0.50 & 0.51 & 0.39 & 0.60 & 0.55 & 1.18  \\
NiHfSn & 6.10 & -0.67 & 0.00 & 0.90 & 0.39 & 0.40 & 0.33 & 0.56 & 0.49 & 1.11  \\
\end{tabular}
\end{ruledtabular}
\end{table*}

To study the effect of the exchange energy we carry out the DFT calculations employing 
the \textsc{QuantumATK} software package ~\cite{QuantumATK,QuantumATKb}. We use linear 
combinations of atomic orbitals (LCAO) as a basis set together with norm-conserving 
PseudoDojo pseudopotentials \cite{VanSetten2018}. For the exchange energy we employ the 
PBE \cite{Perdew1996}, mBJ \cite{TB09}, r$^2$SCAN \cite{SCAN}, HSE06 \cite{HSE06}, and 
PBE0 \cite{PBE0} exchange functionals. The correlation energy in all these cases is 
described by the PW92 correlation functional \cite{PW92}. For determination of the 
ground-state properties of the bulk compounds, we use a $16 \times 16 \times 16$ 
Monkhorst-Pack $\mathbf{k}$-point grid \cite{Monkhorst1976}. We should note that for all the exchange functionals considered, we converged our calculations with the total energy and did not perform just one-shot calculations as sometimes performed in  literature for the mBJ or r$^2$SCAN functionals.

\begin{figure}[htbp]
\begin{center}
\includegraphics[width=0.45\textwidth]{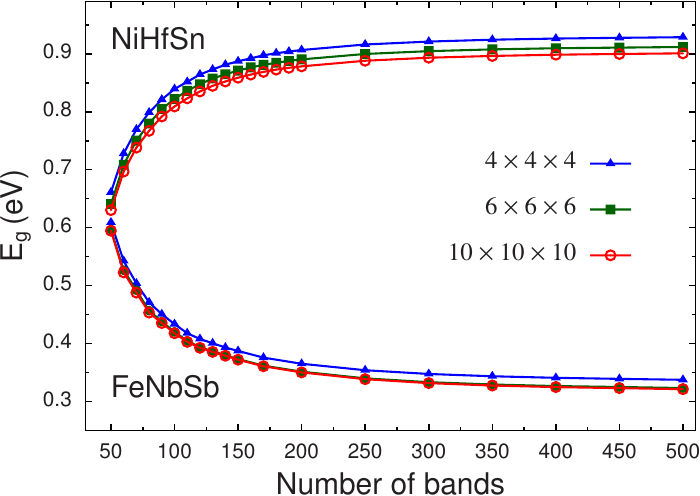}
\end{center}
\vspace*{-0.4cm}
\caption{(Color online) The energy band gap as a function of the number of bands and the 
\textbf{k}-point grid for half-Heusler compounds FeNbSb and NiHfSn.} \label{fig2}
\end{figure}

To study the effect of the correlation energy, we followed the procedure described in detail 
in Ref.\cite{Tas2016}. First, we calculated the ground-state using the FLAPW method as 
implemented in the FLEUR code \cite{FLEUR} within the PBE functional for the exchange and the 
PW92 for the correlation. For all calculations we use angular momentum and plane-wave cutoff 
parameters of $l_{\textrm{max}}=8$ inside the muffin-tin spheres, and 
$k_{\textrm{max}}=4.5$~bohr$^{-1}$ for the outside region. The PBE calculations are performed 
using a $30 \times 30 \times 30$ \textbf{k}-point grid, Then we performed the one-shot $GW$ 
calculations using the SPEX code \cite{SPEX}. In the one-shot $GW$ approach off-diagonal 
elements in the self-energy operator ${\ds \Sigma_\sigma\left(E_{n\bf{k}\sigma}\right)}$ are 
ignored, and corresponding expectation values of the local exchange-correlation potential 
${\ds V_\sigma^{\textrm{XC}}}$ are subtracted in order to prevent double counting. Within this 
framework, the Kohn-Sham (KS) single-particle wavefunctions 
${\ds \varphi_{n\bf{k}\sigma}^{\textrm{KS}}}$ are taken as approximations to the quasiparticle 
(QP) wavefunctions. Hence, the QP energies ${\ds E_{n\bf{k}\sigma}}$ are calculated as a 
first-order perturbation correction to the KS values ${\ds E_{n\bf{k}\sigma}^{\textrm{KS}}}$ 
as, \cite{Aguilera}

\begin{equation}
E_{n\bf{k}\sigma}=E_{n\bf{k}\sigma}^{\textrm{KS}}+\left\langle 
\varphi_{n\bf{k}\sigma}^{\textrm{KS}}| \Sigma_\sigma\left(E_{n\bf{k}\sigma}\right) - 
V_\sigma^{\textrm{XC}}|\varphi_{n\bf{k}\sigma}^{\textrm{KS}}\right\rangle,
\end{equation}
where $n$, \textbf{k}, and $\sigma$ are band index, Bloch vector, and electron spin, 
respectively. The dynamically screened Coulomb interaction $W$ is expanded in the mixed 
product basis set having contributions from the local atom-centered muffin-tin spheres, and 
plane waves in the interstitial region \cite{Kotani}. For the mixed product basis set we used 
the cutoff parameters ${\ds L_{\textrm{max}}=4}$ and ${\ds G_{\textrm{max}}=4}$~bohr$^{-1}$. 
For each compound, a $12 \times 12 \times 12$ \textbf{k}-point grid is used to sample the 
full Brillouin zone. The relativistic corrections are treated at the scalar-relativistic 
level (no spin-orbit coupling) for valence states, while the full Dirac equation is employed 
for the core states. We have converged the band gap with the number of states, and a total of 
500 bands are used for all compounds.
In Fig. \ref{fig2}, we present our convergence tests for the energy band gap of FeNbSn and 
NiHfSn. For this test, three different \textbf{k}-point grids are used to sample the full 
Brillouin zone: $4 \times 4 \times 4$, $6 \times 6 \times 6$, and $10 \times 10 \times 10$. 
We fixed the \textbf{k}-grid and varied the number of bands taken into account for the 
calculations. We observed that the band gap converges to within less than 10 meV for all 
compounds under study when 300 bands are included.

The \textbf{k}-grid has little effect on the results. For a small number of bands, the energy 
band gap is close for both compounds. However, as the number of bands is increased, the values 
for NiHfSn and FeNbSn diverge one from the other and converge towards a constant energy band 
gap value.

It is important first to validate the consistency of our results, as different electronic 
band structure methods treat correlation and exchange differently. To achieve this, we 
provide the computed energy gap ($E_g$) under the PBE approximation using both the FLEUR and 
\textsc{QuantumATK} electronic structure codes in Table \ref{table1}. Our results show a high 
degree of agreement, with differences of less than 0.01 eV observed for all compounds. Thus, 
it can be inferred that the electronic properties, when using the same functional, are 
unaffected by the specific \textit{ab initio} calculation method used.

\begin{figure*}[htbp]
\begin{center}
\includegraphics[width=0.80\textwidth]{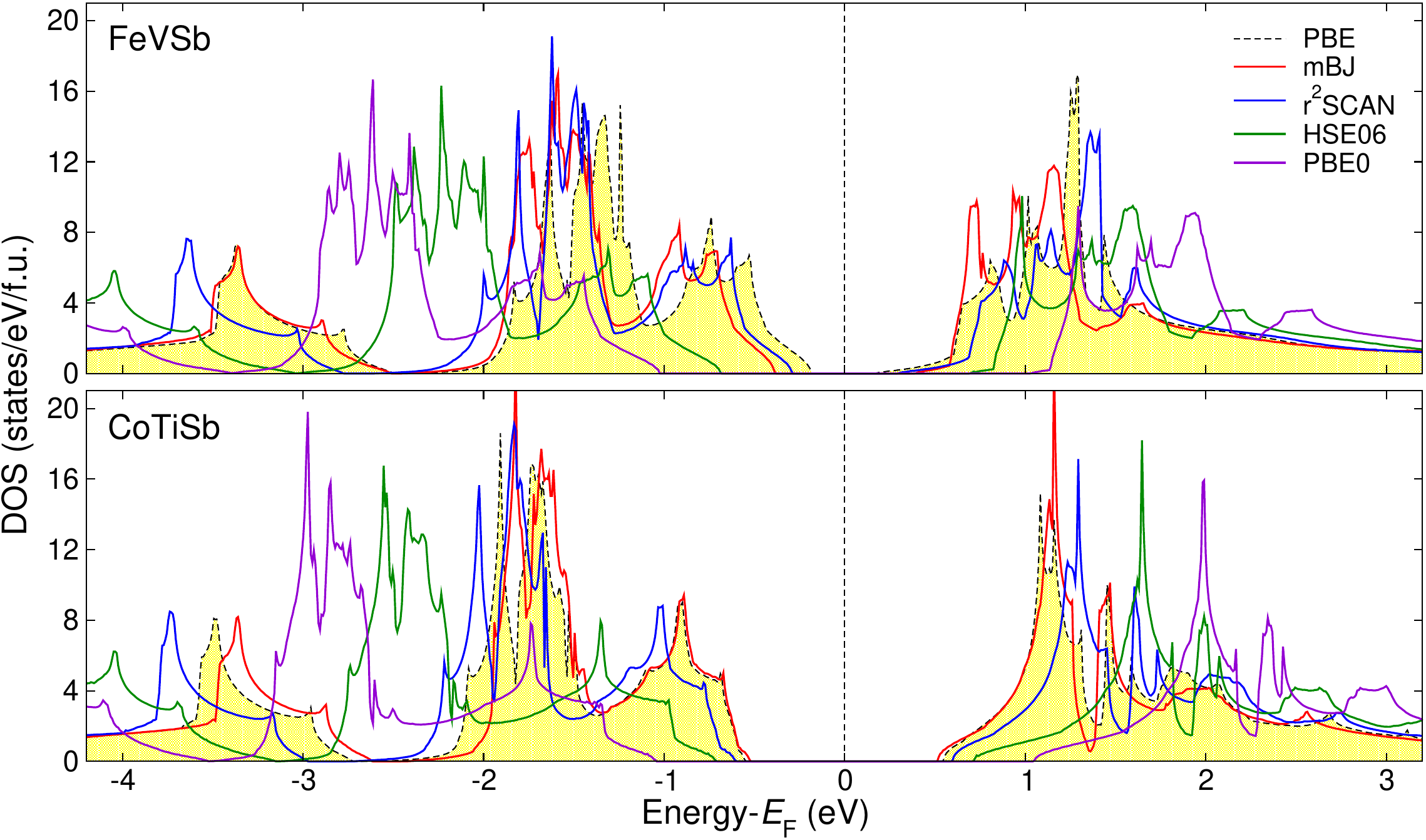}
\end{center}
\vspace*{-0.5cm}
\caption{(Color online) Density of states (DOS) for the half-Heusler compounds FeVSb and 
CoTiSb calculated using the \textsc{QuantumATK} code and various exchange functionals. 
The zero energy corresponds to the Fermi level.} \label{fig3}
\end{figure*}

\section{\label{sec4} The role of exchange}

We begin our discussion by focusing on the role of the exchange energy $E_\mathrm{x}$. For 
this study, we used the \textsc{QuantumATK} electronic band structure code. As discussed 
above, we first performed standard PBE calculations for all 12 compounds. These calculations 
serve as our reference. In Table \ref{table1}, we present the calculated energy band gaps. 
In Fig. \ref{fig3}, we present the total density of states per formula unit (which coincides 
with the unit cell in our case) for FeVSb and CoTiSb. All compounds were found to be indirect 
semiconductors. As shown in Figs. S3-S14 in the Supplementary Material, the maximum of the 
valence band, as calculated by the PBE functional, is located either at the L or $\Gamma$ 
points, while the minimum of the conduction band is always at the X point. The PBE energy 
gap as shown in Table \ref{table1} ranges between 0.34 eV for FeVSb and 1.14 eV for CoHfSb. 
Overall, the compounds containing Co exhibit larger energy gaps than the Fe- and Ni-based 
compounds. The semiconducting behavior of these compounds is also easily observed in the 
DOS in Fig. \ref{fig3}. The Fermi level is set at the middle of the gap as expected for 
semiconductors. The character of the valence and conduction bands near the edges of the gap 
are due to the orbitals of the transition metal atoms as can be derived from the fat-band 
analysis of the PBE calculated electronic band structure in Figs. S1 and S2 of the 
Supplementary Material \cite{Suppl}.

Experimentally, the energy gap has only been determined for CoTiSb in Ref. \cite{Ouardi2012}. 
A value of about 1.0 eV was found, which is close to our PBE-calculated value of 1.09 eV. 
However, we should keep in mind that experimental samples are not perfect, which can lead to 
a smaller experimental energy gap. In Figs. S1 and S2 of the Supplementary Material, we 
present the fat-band analysis for the band structure of several compounds under study. For 
CoTiSb, the maximum of the valence band is almost degenerate, with both the L and $\Gamma$ 
points corresponding to the same maximum valence band energy. The minimum of the conduction 
band is located at the X point. In addition to the energy gap, the optical gap can also be 
determined experimentally. This is the smallest excitation energy while keeping the 
$\mathbf{k}$ vector constant. In Table \ref{table2}, we present the direct transition 
energies (in parentheses the PBE-calculated values). For CoTiSb, the $L \rightarrow L$ 
transition energy corresponds to the optical gap according to PBE and has a value of 1.9 eV, 
which is very close to the experimentally determined optical gap of about 1.8 eV in Ref. 
\cite{Ouardi2012}. It is also worth noting that the $X \rightarrow X$ and $W \rightarrow W$ 
transition energies are also very close to the optical gap.

\begin{figure*}[htbp]
\begin{center}
\includegraphics[width=0.80\textwidth]{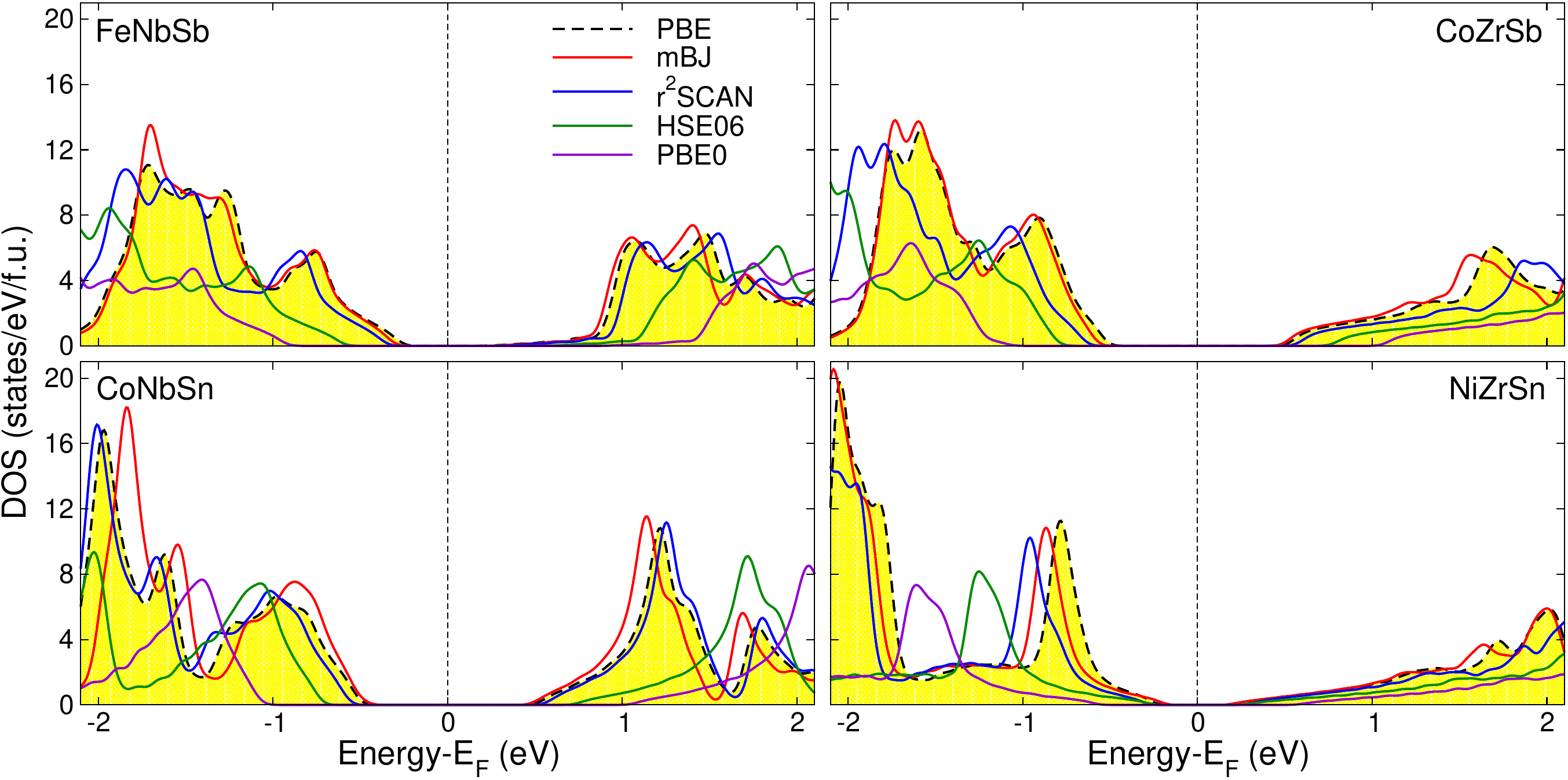}
\end{center}
\vspace*{-0.5cm}
\caption{(Color online) Similar to Fig. \ref{fig3} for the FeNbSb, CoNbSn,  CoZrSb and NiZrSn compounds.}\label{fig4}
\end{figure*}

Using the PBE calculations as a reference, we performed self-consistent calculations using 
both the mBJ and r$^2$SCAN meta-GGA functionals. As we discussed in the introduction, mBJ 
functional is a semi-empirical functional that was developed to produce accurate energy band 
gaps for a few prototype semiconductors \cite{TB09,TB09b}. On the other hand, r$^2$SCAN is a 
complex meta-GGA functional including also the Laplacian of the charge density \cite{r2SCAN}. 
In Table \ref{table1} we present the energy gaps using both functionals. In general, the mBJ 
and r$^2$SCAN energy gaps do not significantly deviate from the PBE values. However, for 
FeVSb, the mBJ functional yields a value that is two times the PBE value. This is because the 
mBJ functional is semi-empirical and does not always have a monotone behavior with respect to 
PBE. In most cases, the mBJ energy gap is larger than the PBE value. However, there are also 
cases, such as the Ni-based compounds, where the mBJ functional gives a smaller energy gap. 
The r$^2$SCAN functional, on the other hand, leads to increased energy gaps with respect to 
PBE in all cases. This is because PBE is well-known to underestimate in general the gaps, and 
SCAN is a considerable improvement in the description of the exchange energy. This behavior 
is reflected in the calculated DOS presented in Fig. \ref{fig3}. For FeVSb, both the mBJ and 
r$^2$SCAN functionals lead to a shift of the valence band maximum towards lower energies, 
widening the energy gap. On the other hand, in the case of CoTiSb, both PBE and mBJ produce a 
similar band structure, especially around the gap. Hence, the PBE and mBJ energy gaps differ 
by only 0.06 eV. In CoTiSb, the r$^2$SCAN functional leads to a small shift of both the 
valence and conduction bands and thus to a small widening of the energy gap of about 10\%.

In the last step, we employed the hybrid HSE06 and PBE0 functionals. Hybrid functionals mix 
the exact Hartree exchange energy and the GGA exchange energy (the PBE functional in our case). 
This is achieved using ad-hoc coefficients, and there is no guarantee that such a scheme will 
work in all cases. The PBE0 functional, which was the initial hybrid functional to be derived 
\cite{PBE0}, leads to energy band gaps that are several times larger than the PBE, mBJ, and 
r$^2$SCAN values as shown in Table \ref{table1}. For example, for CoTiSb, which has an 
experimentally determined energy gap of 1.09 eV, PBE0 gives an energy gap of 2.09 eV, which 
is double the experimental value. The HSE06 functional was developed to speed up the 
convergence of self-consistent calculations, especially in metallic systems \cite{HSE03,HSE06}. 
In general, it leads to energy gaps that are in between the PBE and PBE0 values. This is also 
the case for the twelve compounds under study. HSE06 leads to a smaller shift of the bands in 
the DOS with respect to PBE0 (see Fig. \ref{fig3}). For the Fe-based compounds, HSE06 leads 
to a large increase in the energy gaps, while for the Ni-based compounds the HSE06-calculated 
energy gaps are comparable to the r$^2$SCAN gaps. For the Co-based compounds, HSE06 leads to 
energy gaps that are in general less than 50\% larger than the r$^2$SCAN gaps. 

To elucidate the different behavior for the Ni-based compounds 
we plot in Fig. \ref{fig4} the total DOS for the four 
compounds which contain Nb or Zr atoms. In general the behavior for 
the meta-GGA  functionals is similar to the behavior observed for the  
CoTiSb compound and discussed above. Changes are more pronounced when 
comparing the DOS produced with 
the PBE and the HSE06/PBE0 functionals. In the case of NiZrSn the slope 
of the lower-lying conduction bands is similar for all functionals 
leading to similar energy gaps as mentioned above and the pronounced 
changes in the HSE06/PBE0 DOS are 
observed for higher energies.

To conclude, the GGA and meta-GGA functionals seem to be more adequate for describing the 
exchange energy in CoTiSb, a prototype compound for which experimental data exist. The 
r$^2$SCAN functional has an advantage over mBJ because it is not semi-empirical. The hybrid 
functionals produce too large energy gaps, mainly due to the arbitrary mixing factors of the 
Hartree and GGA exchange energies. This may work for a few semiconductors, but its 
applicability in all cases is questionable.

\begin{table}
\caption{\label{table2} Direct transition energies (in eV) between certain high-symmetry 
points in the Brillouin zone calculated within the $GW$ approximation for all half-Heusler 
compounds studied. The corresponding values obtained with the PBE functional are given in 
parentheses.}
\begin{ruledtabular}
\begin{tabular}{lllll}
 XYZ & ~ $\Gamma \rightarrow \Gamma$ & ~ $\mathrm{X} \rightarrow \mathrm{X}$ & ~ 
 $\mathrm{W} \rightarrow \mathrm{W}$ & ~ $\mathrm{L} \rightarrow \mathrm{L}$ \\
\hline\hline
 FeVSb  & 2.19 (1.98) & 0.94 (0.95) & 1.53 (1.40) & 1.33 (1.21) \\
 FeNbSb & 3.41 (2.82) & 0.75 (1.16) & 2.11 (1.94) & 1.74 (1.78) \\
 FeTaSb & 3.69 (3.03) & 1.12 (1.49) & 2.25 (2.12) & 1.82 (2.04) \\
 CoTiSb & 2.30 (1.85) & 1.86 (1.91) & 2.16 (1.94) & 2.12 (1.90) \\
 CoZrSb & 3.23 (2.59) & 2.10 (1.77) & 2.22 (2.19) & 2.50 (2.59) \\
 CoHfSb & 3.01 (2.55) & 1.79 (1.87) & 2.45 (2.44) & 2.42 (2.79) \\
 CoVSn  & 1.67 (1.49) & 2.04 (1.49) & 1.52 (1.26) & 1.29 (1.22) \\
 CoNbSn & 2.71 (2.22) & 2.19 (1.30) & 1.75 (1.56) & 1.78 (1.86) \\
 CoTaSn & 2.99 (2.43) & 2.42 (1.51) & 2.00 (1.87) & 2.21 (2.22) \\
 NiTiSn & 1.69 (1.39) & 2.10 (1.20) & 2.25 (1.92) & 2.12 (1.88) \\
 NiZrSn & 2.52 (2.06) & 1.67 (0.91) & 2.23 (2.17) & 2.74 (2.70) \\
 NiHfSn & 2.79 (2.29) & 1.91 (1.12) & 2.61 (2.55) & 3.14 (3.05) \\
\end{tabular}
\end{ruledtabular}
\end{table}

\section{\label{sec5} The role of correlation}

As mentioned in the Introduction, current \textit{ab-initio} calculations based on the DFT 
adopted for the description of the correlation energy, $E_\mathrm{c}$, the functional 
developed by Perdew and Wang in 1992 (PW92) irrespective of the nature of the exchange 
functional (LDA, GGA, meta-GGA, or hybrid) \cite{PW92}. A common way to include the 
correlation energy correctly is the LDA(GGA)+U scheme which accounts only for the static 
correlations \cite{LDAU1,LDAU2}. In order to capture also the dynamic correlations one has 
to employ the more elaborated LDA(GGA)+DMFT method \cite{DMFT1,DMFT2}. The latter has been 
shown to change considerably, with respect to LDA(GGA)+$U$, important details of the band 
structure like in the case of VAs \cite{VAs}. The failure of DFT methods to accurately 
describe the correlations arises from the fact that the Kohn-Sham eigenvalues do not have 
a physical meaning. The $GW$ approximation for the electronic self-energy can be used to 
correct this problem, and it can be used to describe correctly the excitation spectra in 
materials, including semiconductors \cite{GW,GW2,GW3}.

We performed $GW$ calculations for all twelve compounds as described in Sec. \ref{sec3}. 
The results are shown in Table \ref{table1}. Although there is a common belief that $GW$ 
should increase values of the energy band gaps with respect to PBE, this is not correct. 
The $GW$ offers a more accurate description of the band structure and thus more accurate 
values of the energy band gap. Whether it increases or decreases the energy band gap with 
respect to PBE depends on PBE itself; there are cases where PBE,and consequently also the 
meta-GGA functionals, overestimates the energy 
band gaps. This explains the behavior of the $GW$ energy band gaps. There are compounds 
like NIZrSn and NiHfSn where the $GW$ band gap is double the corresponding PBE band gap, 
and there are compounds like FeNbSb and FeTaSb where the $GW$ energy band gap is smaller 
than the corresponding PBE band gaps.

To further explain this behavior, we present the direct transition energies obtained using 
both $GW$ and PBE in Table \ref{table2}. For completeness, we also present the indirect 
transition energies in Table S1 of the Supplementary Material \cite{Suppl}. The most 
noticeable difference occurs for the $\Gamma \rightarrow \Gamma$ and 
$\mathrm{L} \rightarrow \mathrm{L}$ transition energies. In general, the former increase 
while the latter decrease when we compare $GW$ with PBE results. This is because the $GW$ 
approximation improves the description of the valence band maximum, which is located at both 
the $\Gamma$ and L points in the PBE case. However, within the $GW$, there is a small shift 
of the valence bands around these points. This can be seen in the band structures presented 
in Figs. S3-S14 of the Supplementary Material \cite{Suppl}. As a result of this shift, the 
indirect gap is not always between the L and X points in the Brillouin zone. In some cases, 
the indirect gap is between the $\Gamma$ and X points.

Overall one can conclude from Figs. S3-S14 of the Supplementary Material that within both PBE and $GW$ the conduction band minimum is always located 
at the X point \cite{Suppl}. For the three Fe(V, Nb or Ta)Sb and the three
Co(V, Nb or Ta)Sn compounds  the valence band maximum is always located at 
the L point, while for the three
Ni(Ti, Zr or Hf)Sn compounds at the $\Gamma$ point. 
The Co(Ti, Zr or Hf) compounds are an interesting case since within PBE the highest valence band energies at the L and $\Gamma$ points are almost identical. 
$GW$ leads to a shift of 
the valence band at the $\Gamma$ point and now the valence band maximum for these three 
compounds is clearly located  at the L point.  

\section{\label{sec6} Summary and conclusions}

\textit{Ab initio} (first-principles) electronic band structure calculations are a powerful 
tool for investigating the properties of materials. In particular, the determination of the 
energy band gap width and optical gap is of paramount importance for semiconductors in a 
variety of applications. In this study, we investigated the influence of exchange and 
correlation energies on the electronic properties of twelve half-Heusler compounds, each with 
18 valence electrons per formula unit. We first examined the impact of the exchange energy 
functional. We considered a variety of functionals, including the conventional 
generalized-gradient approximation (GGA) functional, various meta-GGA functionals, and hybrid 
functionals. Our results revealed significant discrepancies in the calculated energy band gap 
for these half-Heusler compounds, depending on the exchange functional used. Hybrid 
functionals produced band gaps that were larger than those obtained from conventional GGA 
calculations and experimental measurements. On the other hand, non-(semi-)empirical meta-GGA 
functionals, such as r$^2$SCAN, emerged as the optimal choice for probing the electronic 
properties of these compounds. We then used the $GW$ approximation to investigate the effects 
of correlations. Our findings emphasized the importance of correlation effects in determining 
the energy band gap of the half-Heusler compounds. Therefore, our results highlight the need 
for further comparative calculations involving various semiconductors to gain a deeper 
understanding of the behavior of each functional in relation to experimental observations.

\section*{Supplementary Material}
See the supplementary material for additional figures and tables.

\begin{acknowledgments}
This work was supported by SFB CRC/TRR 227 of Deutsche Forschungsgemeinschaft (DFG) and by the 
European Union (EFRE) via Grant No: ZS/2016/06/79307. M. T. acknowledges the TUBITAK ULAKBIM, 
High Performance and Grid Computing Center (TRUBA resources). B. S. acknowledges financial 
support from Swedish Research Council (grant no. 2022-04309). The computations were enabled 
in project SNIC 2021/3-38 by resources provided by the Swedish National Infrastructure for 
Computing (SNIC) at NSC, PDC, and HPC2N partially funded by the Swedish Research Council 
(Grant No. 2018-05973). B.S. acknowledges allocation of supercomputing hours by PRACE DECI-17 
project `Q2Dtopomat' in Eagle supercomputer in Poland and EuroHPC resources in Karolina 
supercomputer in Czech Republic. 
\end{acknowledgments}

\section*{Data Availability Statement}

Data available on request from the authors

\nocite{*}
\providecommand{\noopsort}[1]{}\providecommand{\singleletter}[1]{#1}%


\begin{thebibliography}{57}
\expandafter\ifx\csname natexlab\endcsname\relax\def\natexlab#1{#1}\fi
\expandafter\ifx\csname bibnamefont\endcsname\relax
  \def\bibnamefont#1{#1}\fi
\expandafter\ifx\csname bibfnamefont\endcsname\relax
  \def\bibfnamefont#1{#1}\fi
\expandafter\ifx\csname citenamefont\endcsname\relax
  \def\citenamefont#1{#1}\fi
\expandafter\ifx\csname url\endcsname\relax
  \def\url#1{\texttt{#1}}\fi
\expandafter\ifx\csname urlprefix\endcsname\relax\def\urlprefix{URL }\fi
\providecommand{\bibinfo}[2]{#2}
\providecommand{\eprint}[2][]{\url{#2}}

\bibitem[{\citenamefont{Hohenberg and Kohn}(1964)}]{Hohenberg1964}
\bibinfo{author}{\bibfnamefont{P.}~\bibnamefont{Hohenberg}} \bibnamefont{and}
  \bibinfo{author}{\bibfnamefont{W.}~\bibnamefont{Kohn}},
  \bibinfo{journal}{Phys. Rev.} \textbf{\bibinfo{volume}{136}},
  \bibinfo{pages}{B864} (\bibinfo{year}{1964}).

\bibitem[{\citenamefont{Kohn and SHam}(1965)}]{Kohn1965}
\bibinfo{author}{\bibfnamefont{W.}~\bibnamefont{Kohn}} \bibnamefont{and}
  \bibinfo{author}{\bibfnamefont{L.~J.} \bibnamefont{SHam}},
  \bibinfo{journal}{Phys. Rev.} \textbf{\bibinfo{volume}{140}},
  \bibinfo{pages}{A1133} (\bibinfo{year}{1965}).

\bibitem[{\citenamefont{Perdew and Schmidt}(2001)}]{Perdew01}
\bibinfo{author}{\bibfnamefont{J.~P.} \bibnamefont{Perdew}} \bibnamefont{and}
  \bibinfo{author}{\bibfnamefont{K.}~\bibnamefont{Schmidt}},
  \bibinfo{journal}{AIP Conference Proceedings} \textbf{\bibinfo{volume}{577}},
  \bibinfo{pages}{1} (\bibinfo{year}{2001}).

\bibitem[{\citenamefont{Wang and Perdew}(1991)}]{Perdew1991}
\bibinfo{author}{\bibfnamefont{Y.}~\bibnamefont{Wang}} \bibnamefont{and}
  \bibinfo{author}{\bibfnamefont{J.~P.} \bibnamefont{Perdew}},
  \bibinfo{journal}{Phys. Rev. B} \textbf{\bibinfo{volume}{44}},
  \bibinfo{pages}{13298} (\bibinfo{year}{1991}).

\bibitem[{\citenamefont{Perdew et~al.}(1996)\citenamefont{Perdew, Burke, and
  Ernzerhof}}]{Perdew1996}
\bibinfo{author}{\bibfnamefont{J.~P.} \bibnamefont{Perdew}},
  \bibinfo{author}{\bibfnamefont{K.}~\bibnamefont{Burke}}, \bibnamefont{and}
  \bibinfo{author}{\bibfnamefont{M.}~\bibnamefont{Ernzerhof}},
  \bibinfo{journal}{Phys. Rev. Lett.} \textbf{\bibinfo{volume}{77}},
  \bibinfo{pages}{3865} (\bibinfo{year}{1996}).

\bibitem[{\citenamefont{Sun et~al.}(2015)\citenamefont{Sun, Ruzsinszky, and
  Perdew}}]{SCAN}
\bibinfo{author}{\bibfnamefont{J.}~\bibnamefont{Sun}},
  \bibinfo{author}{\bibfnamefont{A.}~\bibnamefont{Ruzsinszky}},
  \bibnamefont{and} \bibinfo{author}{\bibfnamefont{J.~P.}
  \bibnamefont{Perdew}}, \bibinfo{journal}{Phys. Rev. Lett.}
  \textbf{\bibinfo{volume}{115}}, \bibinfo{pages}{2036402}
  (\bibinfo{year}{2015}).

\bibitem[{\citenamefont{Furness et~al.}(2020)\citenamefont{Furness, Kaplan,
  Ning, Perdew, and Sun}}]{r2SCAN}
\bibinfo{author}{\bibfnamefont{J.~W.} \bibnamefont{Furness}},
  \bibinfo{author}{\bibfnamefont{A.~D.} \bibnamefont{Kaplan}},
  \bibinfo{author}{\bibfnamefont{J.}~\bibnamefont{Ning}},
  \bibinfo{author}{\bibfnamefont{J.~P.} \bibnamefont{Perdew}},
  \bibnamefont{and} \bibinfo{author}{\bibfnamefont{J.}~\bibnamefont{Sun}},
  \bibinfo{journal}{. Phys. Chem. Lett.} \textbf{\bibinfo{volume}{11}},
  \bibinfo{pages}{8208} (\bibinfo{year}{2020}).

\bibitem[{\citenamefont{Becke}(1993)}]{Becke1993}
\bibinfo{author}{\bibfnamefont{A.~D.} \bibnamefont{Becke}},
  \bibinfo{journal}{J. Chem. Phys.} \textbf{\bibinfo{volume}{98}},
  \bibinfo{pages}{1372} (\bibinfo{year}{1993}).

\bibitem[{\citenamefont{Carlo and Barone}(1999)}]{PBE0}
\bibinfo{author}{\bibfnamefont{A.}~\bibnamefont{Carlo}} \bibnamefont{and}
  \bibinfo{author}{\bibfnamefont{V.}~\bibnamefont{Barone}},
  \bibinfo{journal}{J. Chem. Phys.} \textbf{\bibinfo{volume}{110}},
  \bibinfo{pages}{6158} (\bibinfo{year}{1999}).

\bibitem[{\citenamefont{Heyd et~al.}(2003)\citenamefont{Heyd, Scuseria, and
  Ernzerhof}}]{HSE03}
\bibinfo{author}{\bibfnamefont{J.}~\bibnamefont{Heyd}},
  \bibinfo{author}{\bibfnamefont{G.~E.} \bibnamefont{Scuseria}},
  \bibnamefont{and}
  \bibinfo{author}{\bibfnamefont{M.}~\bibnamefont{Ernzerhof}},
  \bibinfo{journal}{J. Chem. Phys.} \textbf{\bibinfo{volume}{118}},
  \bibinfo{pages}{8207} (\bibinfo{year}{2003}).

\bibitem[{\citenamefont{Heyd et~al.}(2006)\citenamefont{Heyd, Scuseria, and
  Ernzerhof}}]{HSE06}
\bibinfo{author}{\bibfnamefont{J.}~\bibnamefont{Heyd}},
  \bibinfo{author}{\bibfnamefont{G.~E.} \bibnamefont{Scuseria}},
  \bibnamefont{and}
  \bibinfo{author}{\bibfnamefont{M.}~\bibnamefont{Ernzerhof}},
  \bibinfo{journal}{J. Chem. Phys.} \textbf{\bibinfo{volume}{124}},
  \bibinfo{pages}{219906} (\bibinfo{year}{2006}).

\bibitem[{\citenamefont{Becke and Johnson}(2006)}]{BJ}
\bibinfo{author}{\bibfnamefont{A.~D.} \bibnamefont{Becke}} \bibnamefont{and}
  \bibinfo{author}{\bibfnamefont{E.~R.} \bibnamefont{Johnson}},
  \bibinfo{journal}{J. Chem. Phys.} \textbf{\bibinfo{volume}{124}},
  \bibinfo{pages}{221101} (\bibinfo{year}{2006}).

\bibitem[{\citenamefont{Tran and Blaha}(2009)}]{TB09}
\bibinfo{author}{\bibfnamefont{F.}~\bibnamefont{Tran}} \bibnamefont{and}
  \bibinfo{author}{\bibfnamefont{P.}~\bibnamefont{Blaha}},
  \bibinfo{journal}{Phys. Rev. Lett.} \textbf{\bibinfo{volume}{102}},
  \bibinfo{pages}{226401} (\bibinfo{year}{2009}).

\bibitem[{\citenamefont{Haas et~al.}(2009)\citenamefont{Haas, Tran, and
  Blaha}}]{TB09b}
\bibinfo{author}{\bibfnamefont{P.}~\bibnamefont{Haas}},
  \bibinfo{author}{\bibfnamefont{F.}~\bibnamefont{Tran}}, \bibnamefont{and}
  \bibinfo{author}{\bibfnamefont{P.}~\bibnamefont{Blaha}},
  \bibinfo{journal}{Phys. Rev. B} \textbf{\bibinfo{volume}{79}},
  \bibinfo{pages}{085104} (\bibinfo{year}{2009}).

\bibitem[{\citenamefont{Meinert}(2013)}]{Meinert2013}
\bibinfo{author}{\bibfnamefont{M.}~\bibnamefont{Meinert}},
  \bibinfo{journal}{Phys. Rev. B} \textbf{\bibinfo{volume}{87}},
  \bibinfo{pages}{045103} (\bibinfo{year}{2013}).

\bibitem[{\citenamefont{Vosko et~al.}(1980)\citenamefont{Vosko, Wilk, and
  Nusair}}]{VWN80}
\bibinfo{author}{\bibfnamefont{S.~H.} \bibnamefont{Vosko}},
  \bibinfo{author}{\bibfnamefont{L.}~\bibnamefont{Wilk}}, \bibnamefont{and}
  \bibinfo{author}{\bibfnamefont{M.}~\bibnamefont{Nusair}},
  \bibinfo{journal}{Can. J. Phys.} \textbf{\bibinfo{volume}{58}},
  \bibinfo{pages}{1200} (\bibinfo{year}{1980}).

\bibitem[{\citenamefont{Ceperley and Alder}(1980)}]{CA}
\bibinfo{author}{\bibfnamefont{D.~M.} \bibnamefont{Ceperley}} \bibnamefont{and}
  \bibinfo{author}{\bibfnamefont{B.~J.} \bibnamefont{Alder}},
  \bibinfo{journal}{Phys. Rev. Lett.} \textbf{\bibinfo{volume}{45}},
  \bibinfo{pages}{566} (\bibinfo{year}{1980}).

\bibitem[{\citenamefont{Perdew and Wang}(1992)}]{PW92}
\bibinfo{author}{\bibfnamefont{J.~P.} \bibnamefont{Perdew}} \bibnamefont{and}
  \bibinfo{author}{\bibfnamefont{Y.}~\bibnamefont{Wang}},
  \bibinfo{journal}{Phys. Rev. B} \textbf{\bibinfo{volume}{45}},
  \bibinfo{pages}{13244} (\bibinfo{year}{1992}).

\bibitem[{\citenamefont{Karlsson et~al.}(2010)\citenamefont{Karlsson,
  Aryasetiawan, and Jepsen}}]{LDAU1}
\bibinfo{author}{\bibfnamefont{K.}~\bibnamefont{Karlsson}},
  \bibinfo{author}{\bibfnamefont{F.}~\bibnamefont{Aryasetiawan}},
  \bibnamefont{and} \bibinfo{author}{\bibfnamefont{O.}~\bibnamefont{Jepsen}},
  \bibinfo{journal}{Phys. Rev. B} \textbf{\bibinfo{volume}{81}},
  \bibinfo{pages}{245113} (\bibinfo{year}{2010}).

\bibitem[{\citenamefont{Solovyev}(2008)}]{LDAU2}
\bibinfo{author}{\bibfnamefont{I.}~\bibnamefont{Solovyev}},
  \bibinfo{journal}{J. Phys.: Condens. Matter} \textbf{\bibinfo{volume}{20}},
  \bibinfo{pages}{293201} (\bibinfo{year}{2008}).

\bibitem[{\citenamefont{Min\'ar}(2011)}]{DMFT1}
\bibinfo{author}{\bibfnamefont{J.}~\bibnamefont{Min\'ar}}, \bibinfo{journal}{J.
  Phys.: Condens. Matter} \textbf{\bibinfo{volume}{23}},
  \bibinfo{pages}{253201} (\bibinfo{year}{2011}).

\bibitem[{\citenamefont{Lechermann et~al.}(2006)\citenamefont{Lechermann,
  Georges, Poteryaev, Biermann, Posternak, Yamasaki, and Andersen}}]{DMFT2}
\bibinfo{author}{\bibfnamefont{F.}~\bibnamefont{Lechermann}},
  \bibinfo{author}{\bibfnamefont{A.}~\bibnamefont{Georges}},
  \bibinfo{author}{\bibfnamefont{A.}~\bibnamefont{Poteryaev}},
  \bibinfo{author}{\bibfnamefont{S.}~\bibnamefont{Biermann}},
  \bibinfo{author}{\bibfnamefont{M.}~\bibnamefont{Posternak}},
  \bibinfo{author}{\bibfnamefont{A.}~\bibnamefont{Yamasaki}}, \bibnamefont{and}
  \bibinfo{author}{\bibfnamefont{O.~K.} \bibnamefont{Andersen}},
  \bibinfo{journal}{Phys. Rev. B} \textbf{\bibinfo{volume}{74}},
  \bibinfo{pages}{125120} (\bibinfo{year}{2006}).

\bibitem[{\citenamefont{{\c{S}}a{\c{s}}{\i}o{\u{g}}lu
  et~al.}(2013)\citenamefont{{\c{S}}a{\c{s}}{\i}o{\u{g}}lu, Galanakis,
  Friedrich, and Bl{\"u}gel}}]{UJ1}
\bibinfo{author}{\bibfnamefont{E.}~\bibnamefont{{\c{S}}a{\c{s}}{\i}o{\u{g}}lu}},
  \bibinfo{author}{\bibfnamefont{I.}~\bibnamefont{Galanakis}},
  \bibinfo{author}{\bibfnamefont{F.}~\bibnamefont{Friedrich}},
  \bibnamefont{and}
  \bibinfo{author}{\bibfnamefont{S.}~\bibnamefont{Bl{\"u}gel}},
  \bibinfo{journal}{Phys. Rev. B} \textbf{\bibinfo{volume}{88}},
  \bibinfo{pages}{134402} (\bibinfo{year}{2013}).

\bibitem[{\citenamefont{Tas et~al.}(2022)\citenamefont{Tas,
  {\c{S}}a{\c{s}}{\i}o{\u{g}}lu, Bl{\"u}gel, Mertig, and Galanakis}}]{UJ2}
\bibinfo{author}{\bibfnamefont{M.}~\bibnamefont{Tas}},
  \bibinfo{author}{\bibfnamefont{E.}~\bibnamefont{{\c{S}}a{\c{s}}{\i}o{\u{g}}lu}},
  \bibinfo{author}{\bibfnamefont{S.}~\bibnamefont{Bl{\"u}gel}},
  \bibinfo{author}{\bibfnamefont{I.}~\bibnamefont{Mertig}}, \bibnamefont{and}
  \bibinfo{author}{\bibfnamefont{I.}~\bibnamefont{Galanakis}},
  \bibinfo{journal}{Phys. Rev. Materials} \textbf{\bibinfo{volume}{6}},
  \bibinfo{pages}{114401} (\bibinfo{year}{2022}).

\bibitem[{\citenamefont{Hedin}(1965)}]{GW}
\bibinfo{author}{\bibfnamefont{L.}~\bibnamefont{Hedin}},
  \bibinfo{journal}{Phys. Rev.} \textbf{\bibinfo{volume}{139}},
  \bibinfo{pages}{A796} (\bibinfo{year}{1965}).

\bibitem[{\citenamefont{Aulbur et~al.}(2000)\citenamefont{Aulbur, Jönsson, and
  W.}}]{GW2}
\bibinfo{author}{\bibfnamefont{W.~G.} \bibnamefont{Aulbur}},
  \bibinfo{author}{\bibfnamefont{L.}~\bibnamefont{Jönsson}}, \bibnamefont{and}
  \bibinfo{author}{\bibfnamefont{W.~J.} \bibnamefont{W.}}
  (\bibinfo{publisher}{Academic Press, New York}, \bibinfo{year}{2000}),
  p.~\bibinfo{pages}{1}.

\bibitem[{\citenamefont{Meinert et~al.}(2012)\citenamefont{Meinert, Friedrich,
  Reiss, and Bl\"ugel}}]{GW3}
\bibinfo{author}{\bibfnamefont{M.}~\bibnamefont{Meinert}},
  \bibinfo{author}{\bibfnamefont{C.}~\bibnamefont{Friedrich}},
  \bibinfo{author}{\bibfnamefont{G.}~\bibnamefont{Reiss}}, \bibnamefont{and}
  \bibinfo{author}{\bibfnamefont{S.}~\bibnamefont{Bl\"ugel}},
  \bibinfo{journal}{Phys. Rev. B} \textbf{\bibinfo{volume}{86}},
  \bibinfo{pages}{245115} (\bibinfo{year}{2012}).

\bibitem[{\citenamefont{Heusler}(1903)}]{Heusler1903}
\bibinfo{author}{\bibfnamefont{F.}~\bibnamefont{Heusler}},
  \bibinfo{journal}{Verh. Dtsch. Phys. Ges.} \textbf{\bibinfo{volume}{12}},
  \bibinfo{pages}{219} (\bibinfo{year}{1903}).

\bibitem[{\citenamefont{Heusler and Take}(1912)}]{Heusler1912}
\bibinfo{author}{\bibfnamefont{F.}~\bibnamefont{Heusler}} \bibnamefont{and}
  \bibinfo{author}{\bibfnamefont{E.}~\bibnamefont{Take}},
  \bibinfo{journal}{Phys. Z.} \textbf{\bibinfo{volume}{13}},
  \bibinfo{pages}{897} (\bibinfo{year}{1912}).

\bibitem[{\citenamefont{Graf et~al.}(2011)\citenamefont{Graf, Felser, and
  Parkin}}]{Graf2011}
\bibinfo{author}{\bibfnamefont{T.}~\bibnamefont{Graf}},
  \bibinfo{author}{\bibfnamefont{C.}~\bibnamefont{Felser}}, \bibnamefont{and}
  \bibinfo{author}{\bibfnamefont{S.~S.~P.} \bibnamefont{Parkin}},
  \bibinfo{journal}{Progr. Sol. St. Chem.} \textbf{\bibinfo{volume}{39}},
  \bibinfo{pages}{1 } (\bibinfo{year}{2011}).

\bibitem[{\citenamefont{Tavares et~al.}(2023)\citenamefont{Tavares, Yang, and
  Meyers}}]{Tavares2023}
\bibinfo{author}{\bibfnamefont{S.}~\bibnamefont{Tavares}},
  \bibinfo{author}{\bibfnamefont{K.}~\bibnamefont{Yang}}, \bibnamefont{and}
  \bibinfo{author}{\bibfnamefont{M.~A.} \bibnamefont{Meyers}},
  \bibinfo{journal}{Progr. Mat. Sci.} \textbf{\bibinfo{volume}{132}},
  \bibinfo{pages}{101017} (\bibinfo{year}{2023}).

\bibitem[{\citenamefont{Chatterjee et~al.}(2022)\citenamefont{Chatterjee,
  Chatterjee, Giri, and Majumdar}}]{Chatterjee2022}
\bibinfo{author}{\bibfnamefont{S.}~\bibnamefont{Chatterjee}},
  \bibinfo{author}{\bibfnamefont{S.}~\bibnamefont{Chatterjee}},
  \bibinfo{author}{\bibfnamefont{S.}~\bibnamefont{Giri}}, \bibnamefont{and}
  \bibinfo{author}{\bibfnamefont{S.}~\bibnamefont{Majumdar}},
  \bibinfo{journal}{J. Phys.: Condens. Matter} \textbf{\bibinfo{volume}{34}},
  \bibinfo{pages}{013001} (\bibinfo{year}{2022}).

\bibitem[{\citenamefont{Galanakis et~al.}(2002)\citenamefont{Galanakis,
  Dederichs, and Papanikolaou}}]{Galanakis2002a}
\bibinfo{author}{\bibfnamefont{I.}~\bibnamefont{Galanakis}},
  \bibinfo{author}{\bibfnamefont{P.}~\bibnamefont{Dederichs}},
  \bibnamefont{and}
  \bibinfo{author}{\bibfnamefont{N.}~\bibnamefont{Papanikolaou}},
  \bibinfo{journal}{Phys. Rev. B} \textbf{\bibinfo{volume}{66}},
  \bibinfo{pages}{134428} (\bibinfo{year}{2002}).

\bibitem[{\citenamefont{Galanakis}(2023)}]{Galanakis2023}
\bibinfo{author}{\bibfnamefont{I.}~\bibnamefont{Galanakis}},
  \bibinfo{journal}{Nanomaterials} \textbf{\bibinfo{volume}{13}}
  (\bibinfo{year}{2023}).

\bibitem[{\citenamefont{Jung et~al.}(2000)\citenamefont{Jung, Koo, and
  Whangbo}}]{Jung2000}
\bibinfo{author}{\bibfnamefont{D.}~\bibnamefont{Jung}},
  \bibinfo{author}{\bibfnamefont{H.-J.} \bibnamefont{Koo}}, \bibnamefont{and}
  \bibinfo{author}{\bibfnamefont{M.-H.} \bibnamefont{Whangbo}},
  \bibinfo{journal}{J. Mol. Struct.:THEOCHEM} \textbf{\bibinfo{volume}{527}},
  \bibinfo{pages}{113} (\bibinfo{year}{2000}).

\bibitem[{\citenamefont{Pierre et~al.}(1994)\citenamefont{Pierre, Skolozdra,
  Gorelenko, and Kouacou}}]{PIERRE1994}
\bibinfo{author}{\bibfnamefont{J.}~\bibnamefont{Pierre}},
  \bibinfo{author}{\bibfnamefont{R.~V.} \bibnamefont{Skolozdra}},
  \bibinfo{author}{\bibfnamefont{Y.~K.} \bibnamefont{Gorelenko}},
  \bibnamefont{and} \bibinfo{author}{\bibfnamefont{M.}~\bibnamefont{Kouacou}},
  \bibinfo{journal}{J. Magn. Magn. Mater.} \textbf{\bibinfo{volume}{134}},
  \bibinfo{pages}{95} (\bibinfo{year}{1994}).

\bibitem[{\citenamefont{Tobola et~al.}(1998)\citenamefont{Tobola, Pierre,
  Kaprzyk, Skolozdra, and Kouacou}}]{Tobola1998}
\bibinfo{author}{\bibfnamefont{J.}~\bibnamefont{Tobola}},
  \bibinfo{author}{\bibfnamefont{J.}~\bibnamefont{Pierre}},
  \bibinfo{author}{\bibfnamefont{S.}~\bibnamefont{Kaprzyk}},
  \bibinfo{author}{\bibfnamefont{R.}~\bibnamefont{Skolozdra}},
  \bibnamefont{and} \bibinfo{author}{\bibfnamefont{M.}~\bibnamefont{Kouacou}},
  \bibinfo{journal}{J. Phys.: Condens. Matter} \textbf{\bibinfo{volume}{10}},
  \bibinfo{pages}{1013} (\bibinfo{year}{1998}).

\bibitem[{\citenamefont{Tobola and Pierre}(2000)}]{Tobola2000}
\bibinfo{author}{\bibfnamefont{J.}~\bibnamefont{Tobola}} \bibnamefont{and}
  \bibinfo{author}{\bibfnamefont{J.}~\bibnamefont{Pierre}},
  \bibinfo{journal}{J. All. Comp.} \textbf{\bibinfo{volume}{296}},
  \bibinfo{pages}{243} (\bibinfo{year}{2000}).

\bibitem[{\citenamefont{Ouardi et~al.}(2012)\citenamefont{Ouardi, Fecher,
  Felser, Schwall, Naghavi, Gloskovskii, Balke, Hamrle, Postava, Pištora
  et~al.}}]{Ouardi2012}
\bibinfo{author}{\bibfnamefont{S.}~\bibnamefont{Ouardi}},
  \bibinfo{author}{\bibfnamefont{G.~H.} \bibnamefont{Fecher}},
  \bibinfo{author}{\bibfnamefont{C.}~\bibnamefont{Felser}},
  \bibinfo{author}{\bibfnamefont{M.}~\bibnamefont{Schwall}},
  \bibinfo{author}{\bibfnamefont{S.~S.} \bibnamefont{Naghavi}},
  \bibinfo{author}{\bibfnamefont{A.}~\bibnamefont{Gloskovskii}},
  \bibinfo{author}{\bibfnamefont{B.}~\bibnamefont{Balke}},
  \bibinfo{author}{\bibfnamefont{J.}~\bibnamefont{Hamrle}},
  \bibinfo{author}{\bibfnamefont{K.}~\bibnamefont{Postava}},
  \bibinfo{author}{\bibfnamefont{J.}~\bibnamefont{Pištora}},
  \bibnamefont{et~al.}, \bibinfo{journal}{Phys. Rev. B}
  \textbf{\bibinfo{volume}{86}}, \bibinfo{pages}{045116}
  (\bibinfo{year}{2012}).

\bibitem[{\citenamefont{Lue et~al.}(2001)\citenamefont{Lue, Oner, Naugle, and
  Ross}}]{Lue2001}
\bibinfo{author}{\bibfnamefont{C.}~\bibnamefont{Lue}},
  \bibinfo{author}{\bibfnamefont{Y.}~\bibnamefont{Oner}},
  \bibinfo{author}{\bibfnamefont{D.}~\bibnamefont{Naugle}}, \bibnamefont{and}
  \bibinfo{author}{\bibfnamefont{J.}~\bibnamefont{Ross}},
  \bibinfo{journal}{IEEE Trans. Magn.} \textbf{\bibinfo{volume}{37}},
  \bibinfo{pages}{2138} (\bibinfo{year}{2001}).

\bibitem[{\citenamefont{Mokhtari et~al.}(2018)\citenamefont{Mokhtari, Dahmane,
  Benabdellah, Zekri, Benalia, and Zekri}}]{Mokhtari2018}
\bibinfo{author}{\bibfnamefont{M.}~\bibnamefont{Mokhtari}},
  \bibinfo{author}{\bibfnamefont{F.}~\bibnamefont{Dahmane}},
  \bibinfo{author}{\bibfnamefont{G.}~\bibnamefont{Benabdellah}},
  \bibinfo{author}{\bibfnamefont{L.}~\bibnamefont{Zekri}},
  \bibinfo{author}{\bibfnamefont{S.}~\bibnamefont{Benalia}}, \bibnamefont{and}
  \bibinfo{author}{\bibfnamefont{N.}~\bibnamefont{Zekri}},
  \bibinfo{journal}{Condens. Matter Phys.} \textbf{\bibinfo{volume}{11}},
  \bibinfo{pages}{43705} (\bibinfo{year}{2018}).

\bibitem[{\citenamefont{Shourov et~al.}(2020)\citenamefont{Shourov, Jacobs,
  Behn, Krebs, Zhang, Strohbeen, Du, Voyles, Brar, Morgan
  et~al.}}]{Shourov2020}
\bibinfo{author}{\bibfnamefont{E.~H.} \bibnamefont{Shourov}},
  \bibinfo{author}{\bibfnamefont{R.}~\bibnamefont{Jacobs}},
  \bibinfo{author}{\bibfnamefont{W.~A.} \bibnamefont{Behn}},
  \bibinfo{author}{\bibfnamefont{Z.~J.} \bibnamefont{Krebs}},
  \bibinfo{author}{\bibfnamefont{C.}~\bibnamefont{Zhang}},
  \bibinfo{author}{\bibfnamefont{P.~J.} \bibnamefont{Strohbeen}},
  \bibinfo{author}{\bibfnamefont{D.}~\bibnamefont{Du}},
  \bibinfo{author}{\bibfnamefont{P.~M.} \bibnamefont{Voyles}},
  \bibinfo{author}{\bibfnamefont{V.~W.} \bibnamefont{Brar}},
  \bibinfo{author}{\bibfnamefont{D.~D.} \bibnamefont{Morgan}},
  \bibnamefont{et~al.}, \bibinfo{journal}{Phys. Rev. Materials}
  \textbf{\bibinfo{volume}{4}}, \bibinfo{pages}{073401} (\bibinfo{year}{2020}).

\bibitem[{\citenamefont{Ma et~al.}(2017)\citenamefont{Ma, Hegde, Munira, Xie,
  Keshavarz, Mildebrath, Wolverton, Ghosh, and Butler}}]{Ma2017}
\bibinfo{author}{\bibfnamefont{J.}~\bibnamefont{Ma}},
  \bibinfo{author}{\bibfnamefont{V.~I.} \bibnamefont{Hegde}},
  \bibinfo{author}{\bibfnamefont{K.}~\bibnamefont{Munira}},
  \bibinfo{author}{\bibfnamefont{Y.}~\bibnamefont{Xie}},
  \bibinfo{author}{\bibfnamefont{S.}~\bibnamefont{Keshavarz}},
  \bibinfo{author}{\bibfnamefont{D.~T.} \bibnamefont{Mildebrath}},
  \bibinfo{author}{\bibfnamefont{C.}~\bibnamefont{Wolverton}},
  \bibinfo{author}{\bibfnamefont{A.~W.} \bibnamefont{Ghosh}}, \bibnamefont{and}
  \bibinfo{author}{\bibfnamefont{W.~H.} \bibnamefont{Butler}},
  \bibinfo{journal}{Phys. Rev. B} \textbf{\bibinfo{volume}{95}},
  \bibinfo{pages}{024411} (\bibinfo{year}{2017}).

\bibitem[{\citenamefont{G\"urb\"uz et~al.}(2023)\citenamefont{G\"urb\"uz,
  Ghosh, \c{S}a\c{s}{\i}o\u{g}lu, Galanakis, Mertig, and Sanyal}}]{Emel2023}
\bibinfo{author}{\bibfnamefont{E.}~\bibnamefont{G\"urb\"uz}},
  \bibinfo{author}{\bibfnamefont{S.}~\bibnamefont{Ghosh}},
  \bibinfo{author}{\bibfnamefont{E.}~\bibnamefont{\c{S}a\c{s}{\i}o\u{g}lu}},
  \bibinfo{author}{\bibfnamefont{I.}~\bibnamefont{Galanakis}},
  \bibinfo{author}{\bibfnamefont{I.}~\bibnamefont{Mertig}}, \bibnamefont{and}
  \bibinfo{author}{\bibfnamefont{B.}~\bibnamefont{Sanyal}},
  \bibinfo{journal}{Phys. Rev. Materials} \textbf{\bibinfo{volume}{7}},
  \bibinfo{pages}{054405} (\bibinfo{year}{2023}).

\bibitem[{oqm()}]{oqmd}
\urlprefix\url{https://oqmd.org/}.

\bibitem[{\citenamefont{Smidstrup et~al.}(2017)\citenamefont{Smidstrup, Stradi,
  Wellendorff, Khomyakov, Vej-Hansen, Lee, Ghosh, J{\'o}nsson, J{\'o}nsson, and
  Stokbro}}]{QuantumATK}
\bibinfo{author}{\bibfnamefont{S.}~\bibnamefont{Smidstrup}},
  \bibinfo{author}{\bibfnamefont{D.}~\bibnamefont{Stradi}},
  \bibinfo{author}{\bibfnamefont{J.}~\bibnamefont{Wellendorff}},
  \bibinfo{author}{\bibfnamefont{P.~A.} \bibnamefont{Khomyakov}},
  \bibinfo{author}{\bibfnamefont{U.~G.} \bibnamefont{Vej-Hansen}},
  \bibinfo{author}{\bibfnamefont{M.-E.} \bibnamefont{Lee}},
  \bibinfo{author}{\bibfnamefont{T.}~\bibnamefont{Ghosh}},
  \bibinfo{author}{\bibfnamefont{E.}~\bibnamefont{J{\'o}nsson}},
  \bibinfo{author}{\bibfnamefont{H.}~\bibnamefont{J{\'o}nsson}},
  \bibnamefont{and} \bibinfo{author}{\bibfnamefont{K.}~\bibnamefont{Stokbro}},
  \bibinfo{journal}{Phys. Rev. B} \textbf{\bibinfo{volume}{96}},
  \bibinfo{pages}{195309} (\bibinfo{year}{2017}).

\bibitem[{\citenamefont{Smidstrup et~al.}(2020)\citenamefont{Smidstrup,
  Markussen, Vancraeyveld, Wellendorff, Schneider, Gunst, Verstichel, Stradi,
  Khomyakov, Vej-Hansen et~al.}}]{QuantumATKb}
\bibinfo{author}{\bibfnamefont{S.}~\bibnamefont{Smidstrup}},
  \bibinfo{author}{\bibfnamefont{T.}~\bibnamefont{Markussen}},
  \bibinfo{author}{\bibfnamefont{P.}~\bibnamefont{Vancraeyveld}},
  \bibinfo{author}{\bibfnamefont{J.}~\bibnamefont{Wellendorff}},
  \bibinfo{author}{\bibfnamefont{J.}~\bibnamefont{Schneider}},
  \bibinfo{author}{\bibfnamefont{T.}~\bibnamefont{Gunst}},
  \bibinfo{author}{\bibfnamefont{B.}~\bibnamefont{Verstichel}},
  \bibinfo{author}{\bibfnamefont{D.}~\bibnamefont{Stradi}},
  \bibinfo{author}{\bibfnamefont{P.~A.} \bibnamefont{Khomyakov}},
  \bibinfo{author}{\bibfnamefont{U.~G.} \bibnamefont{Vej-Hansen}},
  \bibnamefont{et~al.}, \bibinfo{journal}{J. Phys.: Condens. Matter}
  \textbf{\bibinfo{volume}{32}}, \bibinfo{pages}{015901}
  (\bibinfo{year}{2020}).

\bibitem[{\citenamefont{van Setten et~al.}(2018)\citenamefont{van Setten,
  Giantomassi, Bousquet, Verstraete, Hamann, Gonze, and
  Rignanese}}]{VanSetten2018}
\bibinfo{author}{\bibfnamefont{M.~J.} \bibnamefont{van Setten}},
  \bibinfo{author}{\bibfnamefont{M.}~\bibnamefont{Giantomassi}},
  \bibinfo{author}{\bibfnamefont{E.}~\bibnamefont{Bousquet}},
  \bibinfo{author}{\bibfnamefont{M.~J.} \bibnamefont{Verstraete}},
  \bibinfo{author}{\bibfnamefont{D.~R.} \bibnamefont{Hamann}},
  \bibinfo{author}{\bibfnamefont{X.}~\bibnamefont{Gonze}}, \bibnamefont{and}
  \bibinfo{author}{\bibfnamefont{G.~M.} \bibnamefont{Rignanese}},
  \bibinfo{journal}{Comp. Phys. Commun.} \textbf{\bibinfo{volume}{226}},
  \bibinfo{pages}{39} (\bibinfo{year}{2018}).

\bibitem[{\citenamefont{Monkhorst{ } and Pack}(1976)}]{Monkhorst1976}
\bibinfo{author}{\bibfnamefont{H.~J.} \bibnamefont{Monkhorst{ }}}
  \bibnamefont{and} \bibinfo{author}{\bibfnamefont{J.~D.} \bibnamefont{Pack}},
  \bibinfo{journal}{Phys. Rev. B} \textbf{\bibinfo{volume}{13}},
  \bibinfo{pages}{5188} (\bibinfo{year}{1976}).

\bibitem[{\citenamefont{Tas et~al.}(2016)\citenamefont{Tas,
  {\c{S}}a{\c{s}}{\i}o{\u{g}}lu, Galanakis, Friedrich, and
  Bl{\"u}gel}}]{Tas2016}
\bibinfo{author}{\bibfnamefont{M.}~\bibnamefont{Tas}},
  \bibinfo{author}{\bibfnamefont{E.}~\bibnamefont{{\c{S}}a{\c{s}}{\i}o{\u{g}}lu}},
  \bibinfo{author}{\bibfnamefont{I.}~\bibnamefont{Galanakis}},
  \bibinfo{author}{\bibfnamefont{C.}~\bibnamefont{Friedrich}},
  \bibnamefont{and}
  \bibinfo{author}{\bibfnamefont{S.}~\bibnamefont{Bl{\"u}gel}},
  \bibinfo{journal}{Phys. Rev. B} \textbf{\bibinfo{volume}{93}},
  \bibinfo{pages}{195155} (\bibinfo{year}{2016}).

\bibitem[{FLE()}]{FLEUR}
\urlprefix\url{http://www.flapw.de}.

\bibitem[{\citenamefont{Friedrich et~al.}(2010)\citenamefont{Friedrich,
  Bl\"{u}gel, and Schindlmayr}}]{SPEX}
\bibinfo{author}{\bibfnamefont{C.}~\bibnamefont{Friedrich}},
  \bibinfo{author}{\bibfnamefont{S.}~\bibnamefont{Bl\"{u}gel}},
  \bibnamefont{and}
  \bibinfo{author}{\bibfnamefont{A.}~\bibnamefont{Schindlmayr}},
  \bibinfo{journal}{Phys. Rev. B} \textbf{\bibinfo{volume}{81}},
  \bibinfo{pages}{125102} (\bibinfo{year}{2010}).

\bibitem[{\citenamefont{Aguilera et~al.}(2013)\citenamefont{Aguilera,
  Friedrich, Bihlmayer, and Bl\"{u}gel}}]{Aguilera}
\bibinfo{author}{\bibfnamefont{I.}~\bibnamefont{Aguilera}},
  \bibinfo{author}{\bibfnamefont{C.}~\bibnamefont{Friedrich}},
  \bibinfo{author}{\bibfnamefont{G.}~\bibnamefont{Bihlmayer}},
  \bibnamefont{and}
  \bibinfo{author}{\bibfnamefont{S.}~\bibnamefont{Bl\"{u}gel}},
  \bibinfo{journal}{Phys. Rev. B} \textbf{\bibinfo{volume}{88}},
  \bibinfo{pages}{045206} (\bibinfo{year}{2013}).

\bibitem[{\citenamefont{Kotani and van Schilfgaarde}(2002)}]{Kotani}
\bibinfo{author}{\bibfnamefont{T.}~\bibnamefont{Kotani}} \bibnamefont{and}
  \bibinfo{author}{\bibfnamefont{M.}~\bibnamefont{van Schilfgaarde}},
  \bibinfo{journal}{Solid State Commun.} \textbf{\bibinfo{volume}{121}},
  \bibinfo{pages}{461} (\bibinfo{year}{2002}).

\bibitem[{Sup()}]{Suppl}
\bibinfo{note}{Supplemental Material at ........ for more detailed information
  regarding the band structures and the indirect transition energies.}

\bibitem[{\citenamefont{Chioncel et~al.}(2006)\citenamefont{Chioncel,
  Mavropoulos, Lezaic, Bl\"ugel, Arrigoni, Katsnelson, and Lichtenstein}}]{VAs}
\bibinfo{author}{\bibfnamefont{L.}~\bibnamefont{Chioncel}},
  \bibinfo{author}{\bibfnamefont{P.}~\bibnamefont{Mavropoulos}},
  \bibinfo{author}{\bibfnamefont{M.}~\bibnamefont{Lezaic}},
  \bibinfo{author}{\bibfnamefont{S.}~\bibnamefont{Bl\"ugel}},
  \bibinfo{author}{\bibfnamefont{E.}~\bibnamefont{Arrigoni}},
  \bibinfo{author}{\bibfnamefont{M.~I.} \bibnamefont{Katsnelson}},
  \bibnamefont{and} \bibinfo{author}{\bibfnamefont{A.~I.}
  \bibnamefont{Lichtenstein}}, \bibinfo{journal}{Phys. Rev. Lett.}
  \textbf{\bibinfo{volume}{96}}, \bibinfo{pages}{197203}
  (\bibinfo{year}{2006}).

\bibitem[{\citenamefont{Bagayoko}(2014)}]{LDA}
\bibinfo{author}{\bibfnamefont{D.}~\bibnamefont{Bagayoko}},
  \bibinfo{journal}{AIP Advances} \textbf{\bibinfo{volume}{4}},
  \bibinfo{pages}{127104} (\bibinfo{year}{2014}).

\end{thebibliography}
\end{document}